\documentclass[preprint2,longabstract]{aastex}

\def\deg{$^\circ$}
\def\liso{L$_{\rm iso}$}
\def\eiso{E$_{\rm iso}$}
\def\epeak{E$_{\rm peak}$}
\def\t9{T$_{90}$}
\def\tga{t$_{\gamma}$}
\def\tdec{t$_{\rm dec}$}
\def\tb{t$_{\rm b}$}
\def\grb{\object{GRB~110205A}} 

\shorttitle{GRB 110205A}
\shortauthors{Gendre et al.}

\begin{document}

\title{GRB 110205A: Anatomy of a long gamma-ray burst\footnote{Based in part of observations made at the Observatoire de Haute Provence (CNRS), France}}

\author{B. Gendre}
\affil{ASDC, via Galilei Galilei 00044 Frascati (RM) Italy}
\email{bruce.gendre@asdc.asi.it}

\author{J.L. Atteia\altaffilmark{2}}
\affil{Universit\'e de Toulouse; UPS-OMP; IRAP; Toulouse, France}
\altaffiltext{2}{CNRS; IRAP; 14, avenue Edouard Belin, F-31400 Toulouse, France}

\author{M. Bo\"er}
\affil{ARTEMIS (CNRS/UNS/OCA) Observatoire de la C\^ote d'Azur Boulevard de l'Observatoire BP 4229 F-06304 Nice Cedex 4 France}

\author{F. Colas}
\affil{IMCCE, Observatoire de Paris, 77 Avenue Denfert-Rochereau, 75014 Paris, France}

\author{A. Klotz\altaffilmark{2}}
\affil{Universit\'e de Toulouse; UPS-OMP; IRAP; Toulouse, France}

\author{F. Kugel}
\affil{Observatory Chante-Perdrix, Dauban, 04150 Banon, France}

\author{M. Laas-Bourez\altaffilmark{3}}
\affil{University of Western Australia, School of Physics/ICRAR, Crawley, W.A. 6009, Australia}
\altaffiltext{3}{Present address: Observatoire de la cote d'Azur, Grasse, France}

\author{C. Rinner}
\affil{Observatory Chante-Perdrix, Dauban, 04150 Banon, France}

\author{J. Strajnic\altaffilmark{4}}
\affil{Lycee de l'Arc, 84100 Orange, Academie d'Aix-Marseille, France}
\altaffiltext{4}{Observatoire de Haute Provence; CNRS; France}

\author{G. Stratta}
\affil{ASDC, via Galileo Galilei 00044 Frascati (RM) Italy}

\and

\author{F. Vachier}
\affil{IMCCE, Observatoire de Paris, 77 Avenue Denfert-Rochereau, 75014 Paris, France}

\begin{abstract}

The Swift burst \object{GRB 110205A} was a very bright burst visible in the Northern hemisphere. \grb~was intrinsically long and very energetic and it occurred in a low-density interstellar medium environment, leading to delayed afterglow emission and a clear temporal separation
of the main emitting components: prompt emission, reverse shock, and forward shock. Our observations
show several remarkable features of GRB 110205A : the detection of prompt optical emission strongly
correlated with the BAT light curve, with no temporal lag between the two ; the absence of correlation of the X-ray emission compared to the optical and high energy gamma-ray ones during the prompt phase ; and a large optical re-brightening after the end of the prompt phase, that we interpret as a signature of the reverse shock.
Beyond the pedagogical value offered by the excellent multi-wavelength coverage
of a GRB with temporally separated radiating components, we discuss several questions raised
by our observations: the nature of the prompt optical emission and the spectral evolution of the prompt emission
at high-energies (from 0.5 keV to 150 keV) ; the origin of an X-ray flare at the
beginning of the forward shock; and the modeling of the afterglow, including the reverse shock,
in the framework of the classical fireball model. 
\end{abstract}

\maketitle


\clearpage

\section{Introduction}
\label{sec_intro}

Gamma-ray bursts (GRBs), discovered in the late 1960's \citep{kle73}, are the most powerful explosions in the Universe ever since the Big Bang \citep[see e.g.][for reviews]{mes06,ved09}. For about three decades, their exact nature remained elusive. It is only in 1997, due to the efforts to provide a fast re-pointing of the BeppoSAX satellite that observational clues helped to fix their nature \citep[e.g.][]{cos97,van97}. Long GRBs are now thought to be the signature of transient jets from rapidly accreting stellar mass black holes born after the collapse of a massive star in a hypernova \citep{mes06}. 

Since the mid-90, the fireball model \citep{ree92, mes97, pan98} has emerged to explain the GRB phenomenon. This model is based on the ejection of a relativistic fireball/jet during black hole formation, it explains the observed emission by energy dissipation within the fireball (or jet) and when the fireball interacts with the surrounding medium. Observationally, the GRB phenomenon can be divided into various phases. The emission starts with the prompt GRB lasting few to several seconds, which is usually detected at high energies. It is followed by the afterglow, which can be observed at all wavelengths. A consensus exists to attribute the prompt GRB to the internal emission from the jet at distances $\sim 10^8$ km from the source, and the afterglow to the forward shock emission at distances $\sim 10^{12}$ km from the source \citep[see the review of][for more details]{mes06}. One difficulty faced by observers for the interpretation of GRB observations is that the prompt and afterglow emissions are often superimposed in time. While the emission regions of the prompt and of the afterglow are well separated spatially, the relativistic jet travels at the speed of light, and the photons from internal shocks leave the forward region more or less at the same time as the photons from the forward shock.

The main instruments to observe the afterglow quickly are robotic telescopes, that can start observations within seconds of an alert,  and the XRT and UVOT onboard Swift \citep{cas10, klo08b}. This race toward fast re-pointing has allowed various observations of the prompt optical emission by small autonomous telescopes for several GRBs \citep[e.g.][]{ake99, ves06, rac08, klo09c, tho10}. Two general trends have been seen: either a bright optical emission, uncorrelated to the gamma-ray light curve \citep[for 5 to 20\% of GRBs according to the review of ][]{klo09c}, or a faint optical emission correlated with the gamma-ray emission (e.g. \objectname{GRB~050820A}, Vestrand et al. 2006; or \objectname{GRB~081126}, Klotz et al. 2009b). 

In this work, we present a successful observing campaign of \grb~ \citep[T$_{0}$ = 02:02:41 UT,][]{bea11} with facilities ranging from 0.2 to 1.0 meter diameter. The long duration of the gamma--ray emission \citep[T$_{90}$ = $257 \pm 25$ s,][]{mar11} favored early optical observations during the prompt phase. Moreover, the high declination (+67 degrees) means that the field of the GRB is circumpolar for northern observatories. As a consequence, the optical follow-up of \grb~is exceptional. The spectroscopic redshift of \grb~is z = 2.22 \citep{das11,cen11,vre11}. The isotropic equivalent energy radiated by \grb~ during the prompt phase is \eiso = $4.3 \pm 0.4~10^{53}$ erg \citep{sak11,gol11}, and the isotropic equivalent luminosity is \liso = $2 \pm 0.3~10^{52}$ erg \citep{gol11}.

The paper is organized as follows. In Sec. \ref{sec_obs}, we present the data we used. The data reduction procedures and our analysis are explained in Sec. \ref{sec_analysis}. We discuss the prompt phase in Section \ref{sec_prompt}. In the second part of the paper, Sections \ref{sec_ejecta} and \ref{sec_AGmodel}, we discuss the interpretation of the afterglow observations, starting with the late observations and going back in time: first, we discuss the late afterglow, where the explanation has no doubts, and use this modeling to go back in time and constrain the early components: the early afterglow (at X-ray and optical wavelengths), and the reverse shock (at optical wavelengths). We finally conclude in Sec. \ref{sec_conclu}.

Within all this paper, except when indicated, all errors are at the 90\% confidence level, all fits are done with the $\chi^2_\nu$ statistic using Gaussian distributions, all quantities are expressed in the observer frame, and we use a standard lambda-CDM cosmological model (flat Universe, $\Omega_\Lambda = 0.77$) when needed. 

\section{Observations}
\label{sec_obs}

\subsection{Optical data}
\label{ssec_obs_opt}

\paragraph{TAROT Calern observations}

TAROT Calern \citep{klo08} responded promptly to the GCN notice. The first 60 sec. image was trailed using a sampling of 6 sec/pixel. It was obtained in the period from 91.2 to 151.2 s after the trigger. Then, five 30 s images were obtained. A clear filter was used (hereafter C filter). No image was acquired between 378 and 678 s due to a problem of synchronization of the robotic scheduler. Then, long series of 90s to 180s images were obtained up until 13 300 s after the trigger. One third of images was obtained using R filter and the others with the C filter. The TAROT Calern light curve was photometrically calibrated using the R data obtained simultaneously with the T80-OHP and T50-Banon. Because the comparison of optical data and gamma--ray flux is important, a GPS card allow a date accuracy better than 0.1s.

\paragraph{T50-Banon observations}

Images of 180 sec were obtained with the T50-Banon telescope (D=0.50 m, F=1.50 m) of the Observatoire de Chante-Perdrix at Banon, France, through VR filters. The primary focus was equipped with an Sbig STL-11000 (CCD Kodak KAI-11000M front illuminated) and Sbig filters. Magnitudes were calibrated using 11 Loneos calibrated stars in the field of view of NSV 5000. The magnitudes of the GRB optical counterpart were derived for one date, 10 320 s after the trigger, and calibrated using the six stars listed in the Table \ref{Table-x1}.

\paragraph{T80-OHP observations}

The images of 120 sec each were taken at the T80 telescope (D=0.80 m, F=13.3 m) of the Observatoire de Haute-Provence with an Andor 436 (CCD Marconi 47-40 back illuminated) and Johnson-Cousins filters mounted at the Cassegrain focus. BVRI filters were used. Magnitudes were calibrated using 4 Loneos stars in the field of view of RR Boo. At the date of the observation, the elevation of the RR Boo field was the same as the GRB to neglect the airmass corrections. The magnitudes of the GRB optical counterpart were derived for 3 dates between 5040 to 8100 s after the trigger. Moreover, we derived the magnitudes of six stars in the field of view of the GRB (see Table \ref{Table-x1}) to calibrate images obtained with the other telescopes.

\begin{deluxetable}{cccccc}
\tablewidth{0pt}
\tabletypesize{\scriptsize}
\tablecaption{\label{Table-x1} BVRI (Cousin) magnitudes of stars in the field of view of the
optical counterpart of GRB 110205A according the T80-OHP images.}
\tablewidth{0pt}
\tablehead{\colhead{RA (J2000)} & \colhead{DEC(J2000)} & \colhead{V} & \colhead{(B-V)} & \colhead{(V-R)} & \colhead{(V-I)}
}
\startdata
10 58 04.8 & +67 28 35  &              16.59 &  1.29  &        0.73 &  1.26 \\       
10 58 21.0 & +67 29 39  &              15.73 &  0.88  &        0.47 &  0.91 \\         
10 59 02.6 & +67 31 02  &              15.01 &  0.32  &        0.17 &  0.43 \\                
10 58 59.7 & +67 31 11  &              16.04 &  0.81  &        0.43 &  0.84 \\             
10 58 26.1 & +67 33 04  &              14.21 &  0.69  &        0.35 &  0.70 \\               
10 58 25.0 & +67 33 18  &              15.19 &  0.79  &        0.41 &  0.86 \\
\enddata
\end{deluxetable}

\paragraph{T1M Pic du Midi observations}

Several long term follow-up images were taken at the T1M telescope (D=1.05 m, F=12.6 m) of the Observatoire du Pic du Midi. The Nasmyth focus was equipped with an Andor 436 (CCD Marconi 47-40 back illuminated). At t0+ 64 800 s, images of 300 sec were obtained through BVRI filters. We derived an accurate astrometric position of the GRB afterglow:\\
\\
R.A.=10$^h$58$^m$31.14$^s$\\ 
DEC=+67$^\circ$31$\arcmin$30.7$\arcsec$ (J2000.0)\\

Magnitudes were calibrated using stars of the Table \ref{Table-x1}. Late images obtained 1.74, 3.94 and 5.04 days after the trigger were recorded with C filter and were rescaled to the R band using T1M filtered earlier images. This made the T1M observations sensitive down to nearly the 26th magnitude, allowing a strong constraint on the burst geometry, the jet aperture and the derived modeling of the data (see sections 4 and 5).

\paragraph{Other data}

In Tables \ref{Table-non-r} and \ref{Table-r}, we reported magnitudes from French telescopes used for this study. We completed these data with optical and infrared data reported in \citet{cuc11}, and data from GCN circulars \citep{che11,mor11a, mor11b,myu11,sch11,hen11,vol11,ura11a,ura11b,sah11}.

\begin{deluxetable}{cccccc}
\tabletypesize{\scriptsize}
\tablecaption{\label{Table-non-r}Non-R data}
\tablewidth{0pt}
\tablehead{
\colhead{T$_{start}$} & \colhead{T$_{end}$} & \colhead{Filter} & \colhead{Magnitude} & \colhead{Error} & \colhead{Reference}\\
\colhead{(sec)} & \colhead{(sec)} & \colhead{} & \colhead{} & \colhead{} & \colhead{}
}
\startdata
    5045 &     6141 & B &  17.68 &  0.04  & OHP \\
    5045 &     6141 & V &  17.51 &  0.03  & OHP \\
    5045 &     6141 & I &  16.55 &  0.04  & OHP \\
    6563 &     7173 & B &  18.08 &  0.05  & OHP \\
    6563 &     7173 & V &  17.62 &  0.03  & OHP \\
    6563 &     7173 & I &  16.97 &  0.04  & OHP \\
    7428 &     8145 & B &  18.56 &  0.06  & OHP \\
    7428 &     8145 & V &  17.90 &  0.03  & OHP \\
    7428 &     8145 & I &  17.06 &  0.04  & OHP \\
   10695 &    11264 & V &  18.31 &  0.05  & Banon \\
   64602 &    65506 & B &  22.05 &  0.08  & T1M \\
   64602 &    65506 & V &  21.11 &  0.07  & T1M \\
   64602 &    65506 & I &  20.34 &  0.09  & T1M \\
\enddata
\end{deluxetable}

\begin{deluxetable}{ccccc}
\tabletypesize{\scriptsize}
\tablecaption{\label{Table-r}R data}
\tablewidth{0pt}
\tablehead{
\colhead{T$_{start}$} & \colhead{T$_{end}$} &  \colhead{Magnitude} & \colhead{Error} & \colhead{Reference}\\
\colhead{(sec)} & \colhead{(sec)} & \colhead{} & \colhead{} & \colhead{}
}
\startdata
      91 &       92 &  18.54 &  0.68  & TAROT \\
      92 &       98 &  19.50 &  1.23  & TAROT \\
      98 &      104 &  18.12 &  0.50  & TAROT \\
     104 &      110 &  17.99 &  0.46  & TAROT \\
     110 &      116 &  18.07 &  0.49  & TAROT \\
     116 &      122 &  18.60 &  0.71  & TAROT \\
     122 &      128 &  19.37 &  1.14  & TAROT \\
     128 &      134 &  18.82 &  0.82  & TAROT \\
     134 &      140 &  17.63 &  0.35  & TAROT \\
     140 &      146 &  18.58 &  0.69  & TAROT \\
     146 &      151 &  18.54 &  0.68  & TAROT \\
     166 &      196 &  17.84 &  0.08  & TAROT \\
     211 &      241 &  16.90 &  0.08  & TAROT \\
     256 &      286 &  17.61 &  0.08  & TAROT \\
     301 &      331 &  18.09 &  0.08  & TAROT \\
     346 &      376 &  18.37 &  0.08  & TAROT \\
     680 &      770 &  14.87 &  0.08  & TAROT \\
     785 &      875 &  14.35 &  0.08  & TAROT \\
     890 &      980 &  14.12 &  0.08  & TAROT \\
     995 &     1085 &  14.02 &  0.08  & TAROT \\
    1100 &     1190 &  14.20 &  0.08  & TAROT \\
    1205 &     1295 &  14.28 &  0.08  & TAROT \\
    1340 &     1430 &  14.52 &  0.08  & TAROT \\
    1445 &     1535 &  14.53 &  0.08  & TAROT \\
    1550 &     1640 &  14.71 &  0.08  & TAROT \\
    1655 &     1745 &  14.79 &  0.08  & TAROT \\
    1761 &     1851 &  14.97 &  0.08  & TAROT \\
    1865 &     1955 &  15.07 &  0.08  & TAROT \\
    2212 &     2392 &  15.45 &  0.08  & TAROT \\
    2407 &     2587 &  15.66 &  0.08  & TAROT \\
    2602 &     2782 &  15.86 &  0.08  & TAROT \\
    2796 &     2976 &  15.83 &  0.08  & TAROT \\
    2992 &     3172 &  15.98 &  0.08  & TAROT \\
    3186 &     3366 &  16.13 &  0.08  & TAROT \\
    3409 &     3589 &  16.22 &  0.08  & TAROT \\
    3604 &     3784 &  16.14 &  0.08  & TAROT \\
    3798 &     3978 &  16.26 &  0.08  & TAROT \\
    3994 &     4174 &  16.39 &  0.08  & TAROT \\
    4188 &     4368 &  16.45 &  0.08  & TAROT \\
    4383 &     4563 &  16.66 &  0.08  & TAROT \\
    4589 &     4769 &  16.56 &  0.08  & TAROT \\
    4783 &     4963 &  16.56 &  0.08  & TAROT \\
    4979 &     5159 &  16.79 &  0.08  & TAROT \\
    5173 &     5353 &  16.68 &  0.08  & TAROT \\
    5368 &     5548 &  16.76 &  0.08  & TAROT \\
    5045 &     6141 &  16.95 &  0.03  & OHP \\
    5563 &     5743 &  16.94 &  0.08  & TAROT \\
    5785 &     5965 &  16.90 &  0.08  & TAROT \\
    5980 &     6160 &  17.06 &  0.08  & TAROT \\
    6174 &     6354 &  17.09 &  0.08  & TAROT \\
    6369 &     6549 &  17.10 &  0.08  & TAROT \\
    6564 &     6744 &  17.04 &  0.08  & TAROT \\
    6759 &     6939 &  17.21 &  0.08  & TAROT \\
    6563 &     7173 &  17.20 &  0.03  & OHP \\
    6970 &     7150 &  17.35 &  0.08  & TAROT \\
    7166 &     7346 &  17.28 &  0.08  & TAROT \\
    7360 &     7540 &  17.37 &  0.08  & TAROT \\
    7555 &     7735 &  17.53 &  0.08  & TAROT \\
    7428 &     8145 &  17.41 &  0.03  & OHP \\
    7750 &     7930 &  17.51 &  0.08  & TAROT \\
    7945 &     8125 &  17.59 &  0.08  & TAROT \\
    8150 &     8330 &  17.69 &  0.08  & TAROT \\
    8345 &     8525 &  17.72 &  0.08  & TAROT \\
    8539 &     8719 &  17.87 &  0.08  & TAROT \\
    8734 &     8914 &  17.82 &  0.08  & TAROT \\
    8929 &     9109 &  17.77 &  0.08  & TAROT \\
    9123 &     9303 &  17.92 &  0.08  & TAROT \\
    9346 &     9526 &  17.58 &  0.08  & TAROT \\
    9541 &     9721 &  17.98 &  0.08  & TAROT \\
    9736 &     9916 &  18.21 &  0.08  & TAROT \\
    9931 &    10111 &  18.04 &  0.08  & TAROT \\
   10126 &    10306 &  17.78 &  0.15  & TAROT \\
   10107 &    10675 &  17.85 &  0.06  & Banon \\
   10320 &    10500 &  17.87 &  0.15  & TAROT \\
   10525 &    10705 &  17.88 &  0.15  & TAROT \\
   10720 &    10900 &  17.96 &  0.15  & TAROT \\
   10915 &    11095 &  17.96 &  0.15  & TAROT \\
   11110 &    11290 &  17.88 &  0.15  & TAROT \\
   11305 &    11485 &  17.95 &  0.15  & TAROT \\
   11713 &    11893 &  17.98 &  0.15  & TAROT \\
   12298 &    12478 &  18.16 &  0.15  & TAROT \\
   12493 &    12673 &  18.13 &  0.15  & TAROT \\
   13105 &    13285 &  18.25 &  0.15  & TAROT \\
   64602 &    65506 &  20.83 &  0.07  & T1M \\
  146563 &   155125 &  23.15 &  0.31  & T1M \\
  335443 &   345074 &  24.91 &  0.69  & T1M \\
  429094 &   441848 &  25.45 &  0.38  & T1M \\
\enddata

\end{deluxetable}

\subsection{High energy data}
\label{ssec_obs_HE}

\paragraph{BAT}

We retrieved the BAT data of GRB 110205A from the {\em Swift} archive\footnote{http://heasarc.nasa.gov/docs/swift/archive/}. The event file was processed with the latest available calibration files in agreement with the documentation. The task {\it batbinevt} was used to extract spectra and light curves. The spectra were extracted in the same time intervals than the TAROT bins, in order to perform time resolved broad-band spectral studies.

\paragraph{XRT}

We retrieved the XRT data of this event from the {\em Swift} archive. The data were calibrated with the latest available calibration files and screened using the standard filters (i.e. applying good time intervals, grades 0-2 for window timing mode and 0-12 for photon counting mode). The task {\it xselect} was then used to extract spectra and light curves. 

GRB 110205A is a very bright event. The XRT observation strategy can deal with large fluxes, but this burst was so bright that it induced pile-up even in the window timing mode. In addition, the XRT switched to the photon counting mode while the count rate was still high, and thus the initial part of the PC mode observation is also heavily piled-up. We took this into account using the methods listed in \citet{vau06} and \citet{rom06}, and cut the inner part of the extraction region where pile-up is more severe. The task {\it xrtmkarf} was then used to generate the correct ancillary response files for the spectral analysis. We incidentally note that this is not reported in \citet{cuc11}; as we found strong differences in the X-ray data during the prompt phase (see section \ref{sec_prompt}) compared to their work, where the pile-up is the more severe, this fact may explain the discrepancies between this work and \citet{cuc11}. Like for the BAT data, the spectra were extracted in the same time intervals than the TAROT bins, in order to perform time resolved broad-band spectral studies.

\section{Data analysis}
\label{sec_analysis}

\subsection{Temporal binning}
\label{ssec_LAQUELLE}

For clarity, we defined several temporal bins that allow a good referencing when analyzing the data. These bins are listed in Table \ref{table_bin} and are shown in Fig. \ref{fig_lc}. Some of these bins correspond to a single optical data point while others are the sum of several optical points. The last two bins correspond to ''the late afterglow before the final break'' (bin 10) and ''the late afterglow after the final break'' (bin 11).

The optical light curve can be divided into two parts. During the first six temporal bins (corresponding to the first 360 seconds after the trigger) the optical variations seem to be correlated with the gamma-ray flux, we define this part as the early optical light curve. The remaining bins define the late optical light curve.

\begin{deluxetable}{ccc}
\tabletypesize{\scriptsize}
\tablewidth{0pt}
\tablecaption{\label{table_bin} Temporal bins defined for the analysis.}
\tablewidth{0pt}
\tablehead{\colhead{Bin} & \colhead{Start time} & \colhead{End time}\\
\colhead{\#} & \colhead{(s)} & \colhead{(s)}
}
\startdata
 1  &    91.2 &   151.2 \\       
 2  &   165.6 &   195.6 \\       
 3  &   210.6 &   240.6 \\       
 4  &   255.6 &   285.6 \\       
 5  &   301.2 &   331.2 \\       
 6  &   346.2 &   376.2 \\       
 7  &   911   &  1873   \\       
 8  &  5068   &  7651   \\       
 9  & 10842   & 30740   \\       
 10 & 31000   & 61000   \\       
 11 & 61000   &150000   \\            
\enddata
\end{deluxetable}

\subsection{Optical data}
\label{ssec_analysis_opt}

\subsubsection{Photometry methodology}

CCD and filter spectral responses of each instrument are different. It was necessary to calibrate all images in standard system to obtain a composite light curve.

Sextractor \citep{ber96} was used to extract fluxes of stars. Using a filter X (X being R or V), Sextractor gives the flux $F_X$. The conversion between 
fluxes and the magnitudes is given by the following equations:
\begin{equation}
R = Z_R + 2.5 log(F_R) + C_R*(V-R)
\label{equation_1}
\end{equation}
\begin{equation}
V = Z_V + 2.5 log(F_V) + C_V*(V-R)
\label{equation_2}
\end{equation}

$Z_R$, $Z_V$, $C_R$ and $C_V$ are calculated with stars of known V and R magnitudes. To determine these coefficients, we used the Loneos catalog published as "UBVRI photometry of faint field stars" \citep{ski07}. Loneos is based on Johnson-Cousins UBVRI photometry. As a consequence, the R, I colors are calculated in the Cousins system.


\subsubsection{The early optical light curve}

\begin{figure*}
\epsscale{2.0}
\plotone{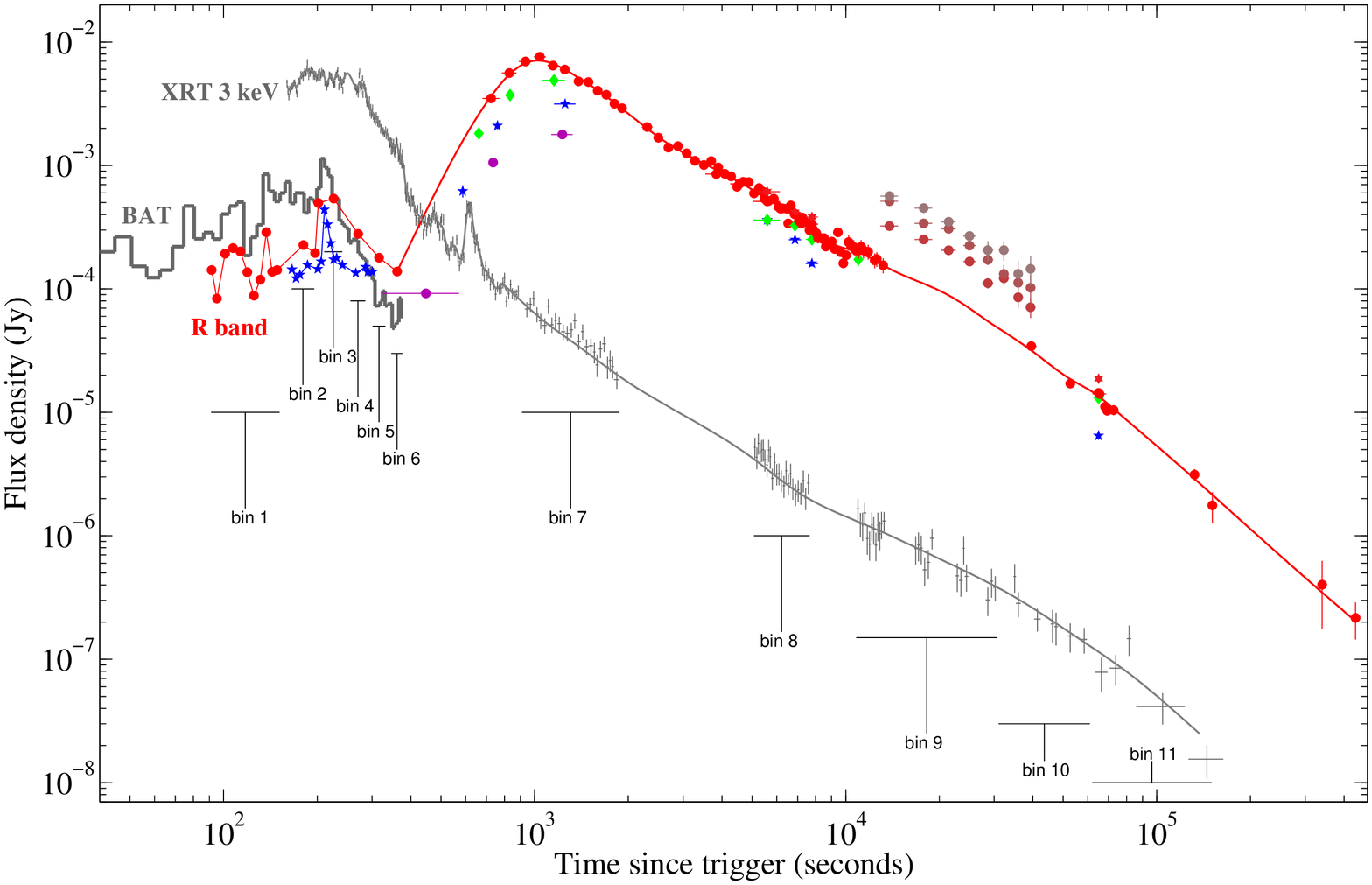}
\caption{\label{fig_lc} Panchromatic light curve of GRB 110205A. The BAT data are indicated as a light gray continuous line. The XRT data are indicated by small plus symbols (with errors). The optical data are indicated by purple circles (U band), blue stars (B band), green diamonds (V band), red circles (R band), and red stars (I band). The JHK data during bin 9 are indicated as brown circles. The red (optical) and dark gray (X-ray) lines are the best fit decay laws (see text for details). See the electronic version for colors.}
\end{figure*}

\begin{figure}
\epsscale{.95}
\plotone{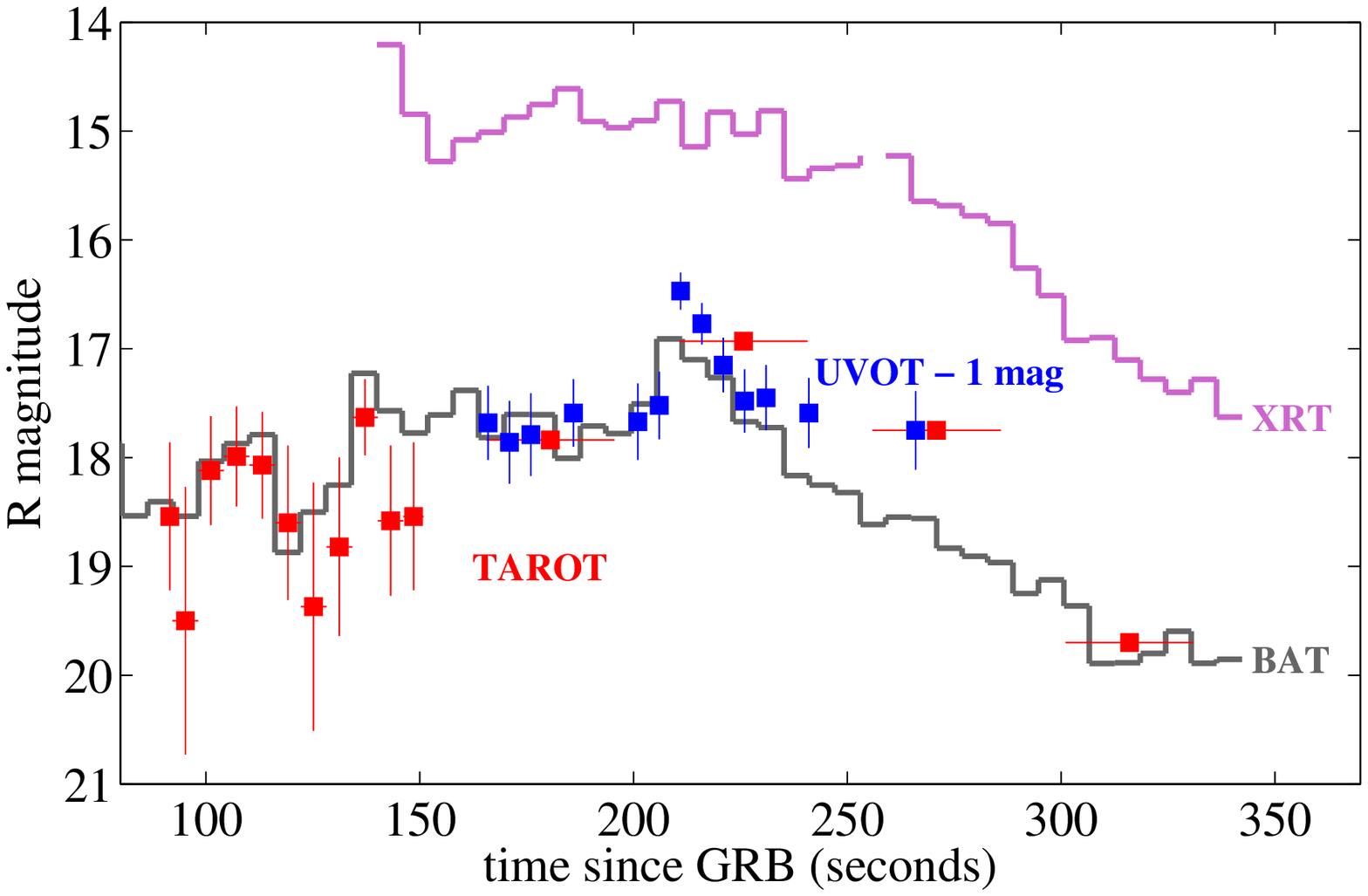}
\caption{\label{fig_prompt} Comparison of the optical, XRT and BAT light curves during the prompt phase. We corrected the optical data for the underlying re-brightening.}
\end{figure}

Figure \ref{fig_lc} shows a striking correlation of optical and gamma--ray fluxes during the prompt phase. A peak of maximum flux occurred about 210 s after the trigger. After the peak, the gamma-ray flux decay is steeper than the optical one. We attribute this fact to the growing contribution of the optical re-brightening that culminates later; indeed, subtracting the rising contribution of the optical re-brightening (extrapolated backward) gives an early optical light curve that closely follows the high-energy one (see Fig. \ref{fig_prompt}). 

\subsubsection{The late optical light curve fit and color indexes}
\label{ssec_analysis_late}
After 360 s, the optical flux rises and reaches a maximum 1014 s after the trigger. Then, the flux decreases continuously until the last observation obtained 5 days after the trigger. We fit the R band light curve using a smoothing spline fitting curve described by \citet{rei67} with a smooth parameter value s=80. We find evidence of small variations during the decay phase (see Fig. \ref{fig_lc}). 

\begin{figure}
\epsscale{.95}
\plotone{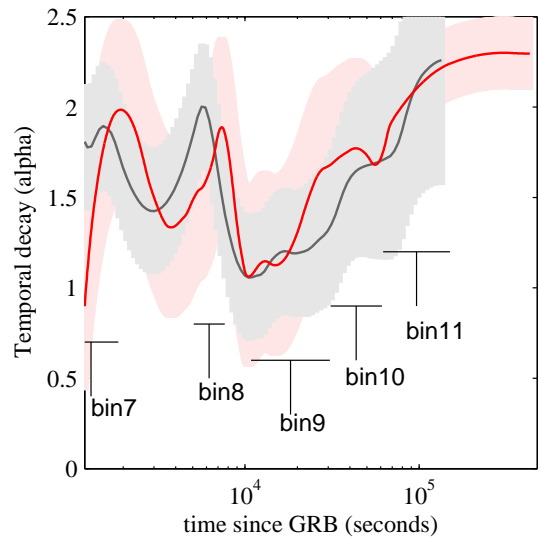}
\caption{\label{fig_alpha}Temporal decay indices of the optical (red line) and X-ray (gray line) light curves in the afterglow phase.}
\end{figure}

From the fit light curve, we computed the temporal decay alpha ($F \propto t^{-\alpha}$). Alpha decrease from -5 to 0 until the maximum of light. Figure \ref{fig_alpha} shows the evolution of the decay index after the re-brightening maximum. The 2 $\sigma$ uncertainty is represented by the shaded area. The continuous evaluation of the decay index is obtained following the method of \citet{rei67}. As one can see, the value is fluctuating, with the presence of a possible plateau at 10 000 s. The mean decay indices are $\alpha=1.5 \pm 0.5$ from 1014 to 61000 s after the trigger, and $\alpha=2.2 \pm 0.2$ after. Due to small erratic variations of the light curve, the break time is not well determined with $t_b = 61 000 \pm 40 000$ s after the trigger.

The light curve shows an achromatic behavior when we plot UBVIJHK data against the R data. The colors are reported in Table \ref{table_color}.

\begin{deluxetable}{ccc}
\tabletypesize{\scriptsize}
\tablecaption{\label{table_color} Color of the afterglow during temporal bin 9.}
\tablewidth{0pt}
\tablehead{\colhead{Color} & \colhead{Value} & \colhead{Error}\\
}
\startdata
$U-R$ & $+0.80$ & 0.11\\
$B-R$ & $+1.00$ & 0.12\\
$V-R$ & $+0.46$ & 0.14\\
$R-I$ & $+0.45$ & 0.15\\
$R-J$ & $+1.60$ & 0.13\\
$R-H$ & $+2.48$ & 0.13\\
$R-K$ & $+3.20$ & 0.13\\
\enddata
\end{deluxetable}

\subsection{High energy data}
\label{ssec_analysis_HE}

\subsubsection{The X-ray light curve}

Figure \ref{fig_prompt} indicates a clear lack of correlation between the early X-ray light curve and the other two bands. It is clear that the optical and gamma-ray peak observed in the light curves is missing in X-ray. Both bands decay after the maximum of this peak while the X-ray light curve remains steady for more than 50 seconds. This is a first hint that something different is happening in X-ray compared to the other bands.

The late X-ray light curve has been fit with the same method than the optical one for easy comparison. As can be seen in Fig. \ref{fig_alpha}, the decay index of the X-ray light curve follows the optical one, with a difference that does not exceed 0.25. After $10^5$\ s, the two light curves have the same temporal variation and reach the same final value $\alpha = 2.2$. We finally note that the decay index changes from $\sim$1.2 to $\sim$2.2 between bin 9 and bin 11.

\subsubsection{High-energy spectra}
\label{ssec_analysis_HEspec}

High-energy  spectra were fit with {\it Xspec} version 12.6.0 \citep{arn96}. We ignored data below 0.3 keV for the XRT data, below 15.0 keV and above 150.0 keV for the BAT instruments. During the prompt phase, both XRT and BAT recorded spectra (except for bin 1, where only BAT data are available). We started fitting both of them separately with single power laws (PL), and found that \textit{(i)} strong spectral variability is present during the prompt phase and the initial afterglow phase, and \textit{(ii)} for temporal bins 2--5 a spectral break is required between the XRT and BAT energy ranges. 

We then used temporal bin 2 to define the spectral model to be used for further analysis, using both XRT and BAT data. We first tried the model used by \citet{sak11}. \citet{sak11} perform a joint fit of the gamma-ray spectrum measured by Swift/BAT (15-150 keV) and Suzaku/WAM (100-3000 keV) with  a power law with exponential cutoff model, and find $\alpha$ = 1.59 (-0.06/+0.07), and \epeak\ = 230 (-65/+135) keV (in the following we use the notation $F_\nu \propto \nu^{-\alpha}$). This shows that \epeak\ is above the energy range of the BAT ; this is also the conclusion of \citet{gol11} who find \epeak\ = 222 keV. Because this value is above the BAT range, we used a simple power law model to fit the XRT and BAT data, completed by an extragalactic absorption component let free to vary at the GRB redshift and a galactic absorption component fixed to the galactic value \citep[$1.61 \times 10^{20}$ cm$^{-2}$][]{dic90}. This model is rejected with a large reduced $\chi^2_\nu$ ($\chi^2_\nu = 2.60$, 265 d.o.f.). The poorness  of the fit shows that a model that correctly fits the spectrum at high energies cannot be extrapolated to the energy range of the XRT, without introducing an additional spectral break in the energy range covered by XRT and BAT. 

Then, we checked whether the data could be fit with an absorbed broken power law. This fit provides a significant improvement over a simple power law, with $\chi^2_\nu = 1.19$, 263 d.o.f. The residuals of the broken PL fit still show a systematic trend (with an excess at low energy) suggesting that the break is too sharp. Considering that the synchrotron emission is made of segments of power law, with no indication of the sharpness of the breaks, we tried to simulate a smooth break using a double broken power law model (this is valid only if the transition between the different segments remains in a small energy range). This improves the fit significantly, leading to $\chi^2_\nu = 1.12$, 261 d.o.f. (FTest null hypothesis probability of $1.3 \times 10^{-4}$). This fit gives a low energy spectral slope $\alpha_1 = -0.35 \pm 0.14$ below 2.8 keV and a spectral slope $\alpha_2 = 0.87 \pm 0.06$ above 9 keV. The transition between these two power laws occurs in a small energy range, thus validating our hypothesis of a smooth break. It is tempting to attribute this smooth transition to one of the characteristic frequencies $\nu_m$ or $\nu_c$. During the prompt phase GRBs must be in the fast cooling regime with $\nu_m > \nu_c$, and the spectral slope below $\nu_c$ is expected to be $\alpha_1 = -0.33$. 

We have thus tried to fit jointly the XRT and the BAT data during the prompt phase (bins 1 to 6) with a double broken power law and the low energy spectral index frozen at -0.33. We assumed that the extragalactic absorption component does not vary during the whole observation (a separate fit to the XRT data alone indicates this hypothesis to be correct within the errors of the fit). This model gives an acceptable fit ($\chi^2_\nu = 1.07$, 1064 d.o.f.), and would explain the non-correlation of the X-ray and gamma-ray bands as a variability of the break energies; we report its results in Table \ref{table_fit_x}.

Incidentally, we note that this burst would have been classified as an X-ray Rich burst because of its hardness ratio.

Last, starting from bin 7, we do not have BAT data anymore, and used a simple power law absorbed by our galaxy and the host galaxy. The fit is good ($\chi^2_\nu = 1.00$, 79 d.o.f.); we report these results in Table \ref{table_fit_x}.

\begin{deluxetable}{ccccccc}
\tabletypesize{\scriptsize}
\tablecaption{\label{table_fit_x}Result of the spectral analysis.}
\tablewidth{0pt}
\tablehead{
\colhead{Segment} & \colhead{Extragalactic} & \colhead{Single}     & \multicolumn{4}{c}{Double broken power law}\\
\colhead{}        & \colhead{N$_H$}         & \colhead{power law}  & \colhead{Break 1} & \colhead{intermediate} & \colhead{Break 2} & \colhead{High-energy}\\
\colhead{}& \colhead{($10^{22}$ cm$^{-2}$)} & \colhead{spectral}   & \colhead{energy}  & \colhead{spectral}     & \colhead{energy}  & \colhead{spectral}\\
\colhead{}        & \colhead{}              & \colhead{index}      & \colhead{(keV)}   & \colhead{index}        & \colhead{(keV)}   & \colhead{index}\\
}
\startdata
1                 & ($1.0 \pm 0.2$)         &   ---                &  [2]              &   [0.0]                &  [5]              & ($0.83 \pm 0.03$)  \\
2                 & ($1.0 \pm 0.2$)         &   ---                & $2.9 \pm 0.5$     & $0.2 \pm 0.2$         & $8.3^{+1.5}_{-1.1}$ & ($0.83 \pm 0.03$) \\
3                 & ($1.0 \pm 0.2$)         &   ---                & $3.0 \pm 0.3$     & $0.42 \pm 0.04$       & $39 \pm 10$         & ($0.83 \pm 0.03$) \\
4                 & ($1.0 \pm 0.2$)         &   ---                & $1.1 \pm 0.2$     & $0.3 \pm 0.1$         & $4.0^{+0.8}_{-0.7}$ & ($0.83 \pm 0.03$) \\
5                 & ($1.0 \pm 0.2$)         &   ---                & $0.6 \pm 0.2$     & $0.58 \pm 0.07$        & $6^{+8}_{-3}$     & ($0.83 \pm 0.03$) \\
6                 & ($1.0 \pm 0.2$)         &   ---                & $<0.60$           & $0.67 \pm 0.06$        &   [10]            & ($0.83 \pm 0.03$)  \\
\hline
7                 & ($<0.48$)               & $0.7 \pm 0.2$        &   ---             &   ---                  &   ---             &  --- \\
8                 & ($<0.48$)               & $1.0 \pm 0.2$        &   ---             &   ---                  &   ---             &  --- \\
9                 & ($<0.48$)               & $1.2 \pm 0.2$        &   ---             &   ---                  &   ---             &  --- \\
\enddata
\tablecomments{We are reporting the energy spectral indices. The temporal segments are defined in Sec \ref{ssec_LAQUELLE}. The low energy double broken power law spectral index is fixed to -0.33 for all segments. Segments 1-6 and 7-9 were fit altogether. Number between parentheses have been tied together during the fit, numbers between square parentheses have been fixed. The low energy and intermediate energy parameters of the first segment are unknown because no XRT data are available at that time: they were fixed to ad hoc values. During segment 6, XRT and BAT are fitted by single power laws with compatible indices and the break2 is not required; we have fixed it to 10.0 keV for fitting convergence.}
\end{deluxetable}

With \eiso\ $\sim 4 \times 10^{53}$ erg, and L$_{\rm iso} \sim 2 \times 10^{52}$ erg s$^{-1}$, \grb\ is a bright burst, but not exceptional. We have checked that it follows the E$_p$ - \eiso\ relation \citep{ama09} and the E$_p$ - L$_{iso}$ relation \citep{yon10}, stressing that we are dealing with a standard GRB.

\subsection{Colors and SED}
\label{ssec_analysis_sed}

We extracted the optical Spectral Energy Distribution (SED) in filters BVR and UBVRIJHK for the temporal bins 7 and 9 respectively, and added the X-ray information using the 2-10 keV flux spectra extracted at the same mean epochs. We corrected the optical data only for the Galactic dust extinction towards the direction of this burst \citep[E(B-V)=0.015,][]{sch98}. In X-ray, the absorption was fixed to the value obtained from the spectral analysis (see previous section and Table \ref{table_fit_x}) and the flux corrected accordingly.

\begin{figure*}
\epsscale{1.80}
\plottwo{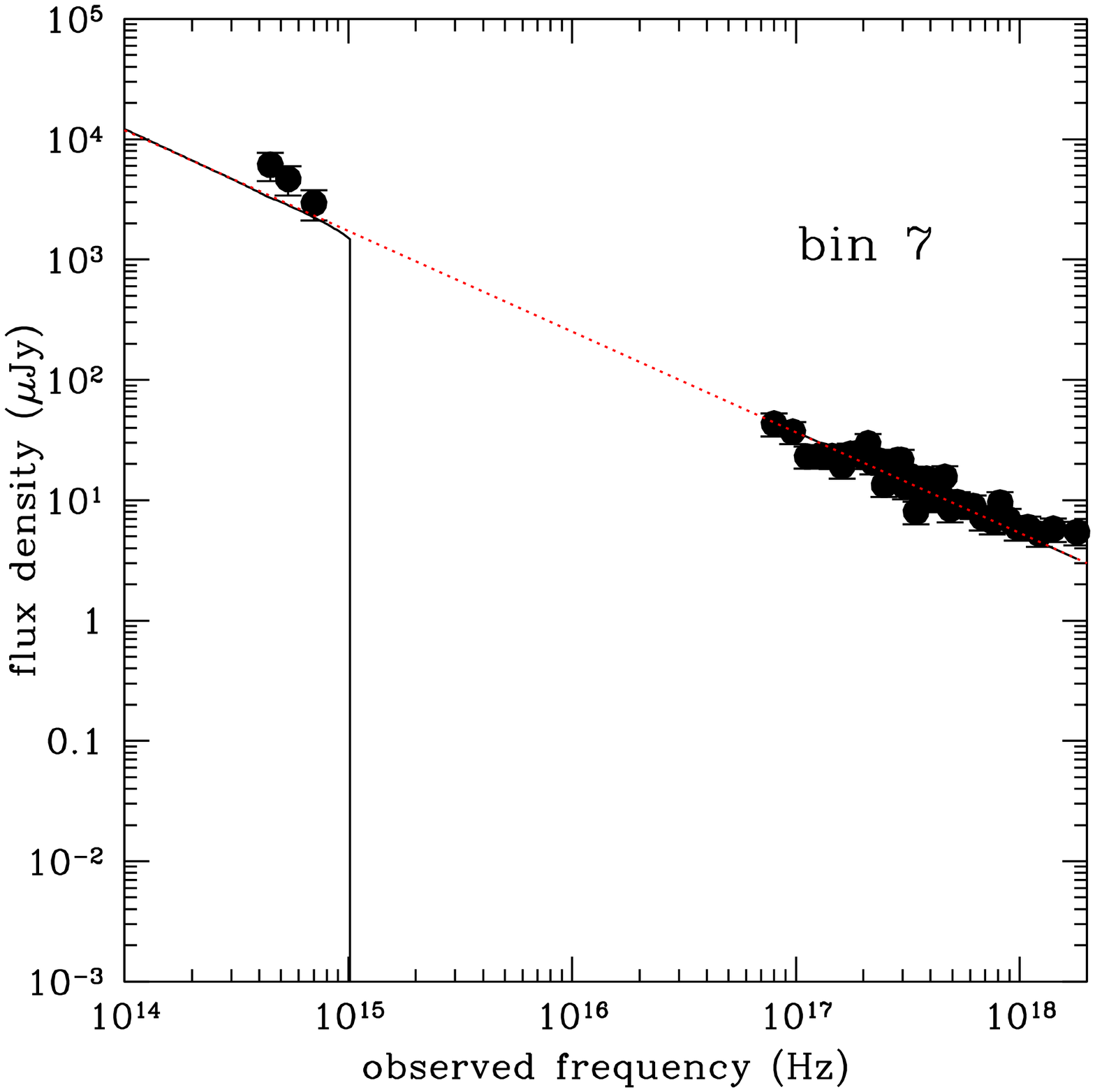}{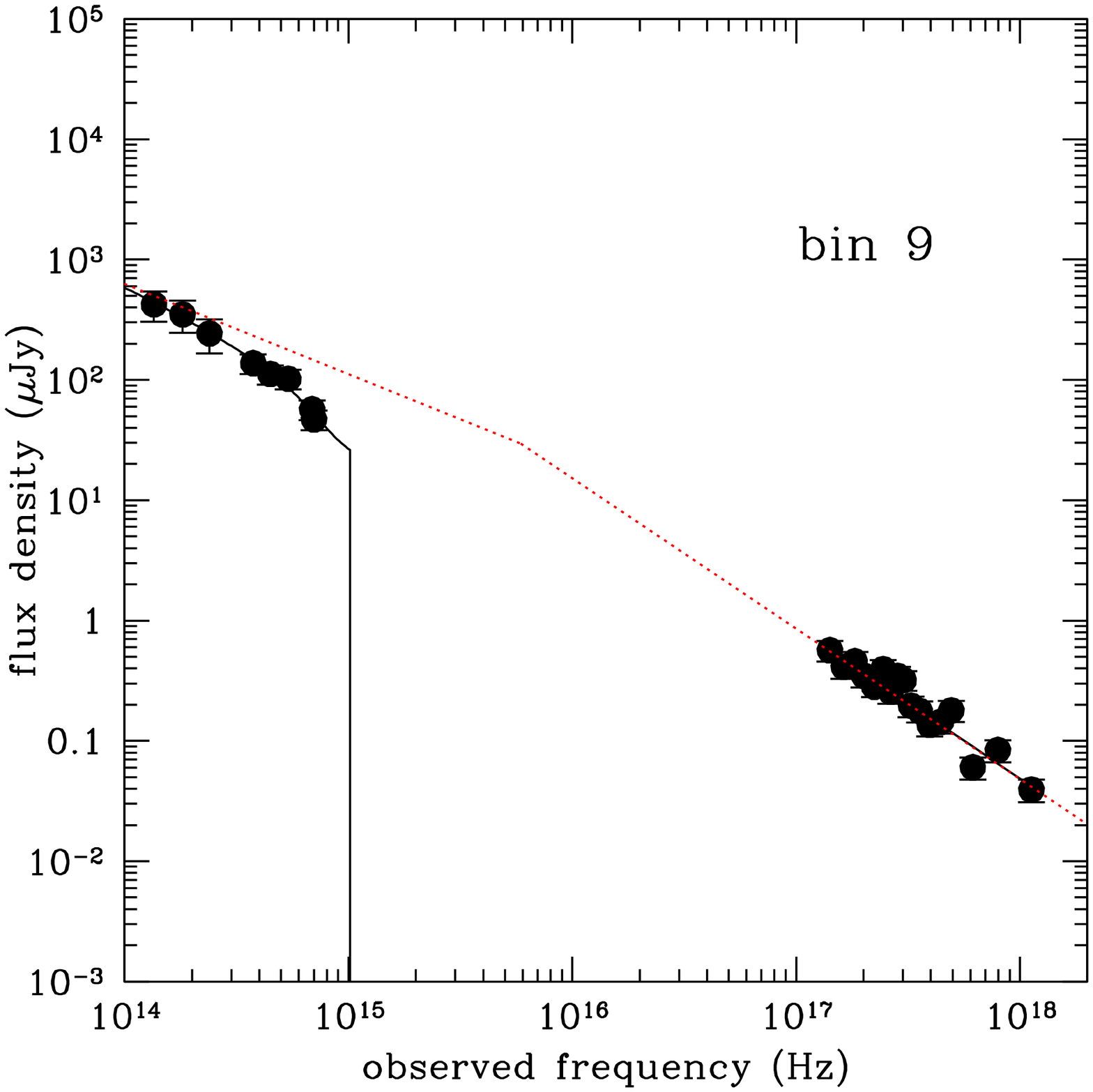}
\caption{\label{fig_giulia1} SED extracted during temporal bins 7 and 9. The red dashed lines indicate the best fit power law model based on X-ray data alone. The solid black line indicate the best fit model based on X-ray and optical data, taking into account optical extinction. See Sect. \ref{ssec_analysis_sed} for fit details.}
\end{figure*}

We first assumed a simple power law model with the spectral index fixed at the X-ray best fit value (allowed to vary within 1 $\sigma$ only), and a Milky Way (MW hereafter) or a Small Magellanic Cloud (SMC) -like dust extinction component. We find for both bin 7 and 9 a null rest frame visual extinction. The best fit spectral index is $\beta_{ox}=0.84 \pm 0.04$ ($\chi_{\nu}^2$=1.23, 39 d.o.f.) and $\beta_{ox} > 0.9$ ($\chi_{\nu}^2=2.9$, 22 d.o.f.) for bins 7 and 9 respectively. Removing the constraint on the spectral index parameter (i.e. letting it free to vary), the fit to bin 7 does not improve, regardless of the dust extinction laws used. In fact, this model under predicts the optical flux (see Fig. \ref{fig_giulia1}, left panel). On the other hand, relaxing the same constraint in bin 9 improves the fit, and the model can marginally fit the data. In that latter case, the best fit parameters are $\beta_{ox} = 1.03 \pm 0.10$ (marginally compatible with the X-ray), and a rest frame visual dust extinction A$_{V, rest}$ of $0.27 \pm 0.10$ mag (MW hypothesis) or $0.14 \pm 0.10$ mag (SMC hypothesis). The reduced chi square remains however high: $\chi_{\nu}^2 = 1.4$, 22 d.o.f.

We then tried a broken power law model, fixing the high energy spectral index to the best fit X-ray value, but allowing now a spectral break between optical and X-ray (with the low energy spectral index value fixed to $\beta_o = \beta_X - 0.5$). For the bin 7 SED, this model cannot fit the data, with a reduced $\chi^2_\nu$ of 3.2 with 39 degrees of freedom. For the bin 9 data set, on the contrary, we obtain a good description of the data, with some improvements assuming a SMC rather than a MW extinction curve. The best fit rest frame visual dust extinction is $0.19 \pm 0.10$ mag, while the spectral break is found at $(5.4 \pm 2.6) \times 10^{15}$ Hz (0.016-0.033 keV; $\chi_{\nu}^2$ = 0.91, 22 d.o.f., see Fig. \ref{fig_giulia1} right panel).

\section{The prompt emission}
\label{sec_prompt}

In this section we discuss the interpretation of the main observational features of the prompt emission of \grb\ within the simple framework of the internal/external shock model with synchrotron emission: 
\begin{itemize}
\item The existence of two spectral breaks: a smooth one at few keV and a high-energy one $\sim 220$ keV (see Sec. \ref{ssec_analysis_HEspec})
\item The good correlation between visible and gamma-ray light-curves
\item The lack of correlation between X-ray and gamma-ray light-curves
\item The overall SED, from visible to gamma-rays
\end{itemize}

\begin{figure*}
\epsscale{2.20}
\plotone{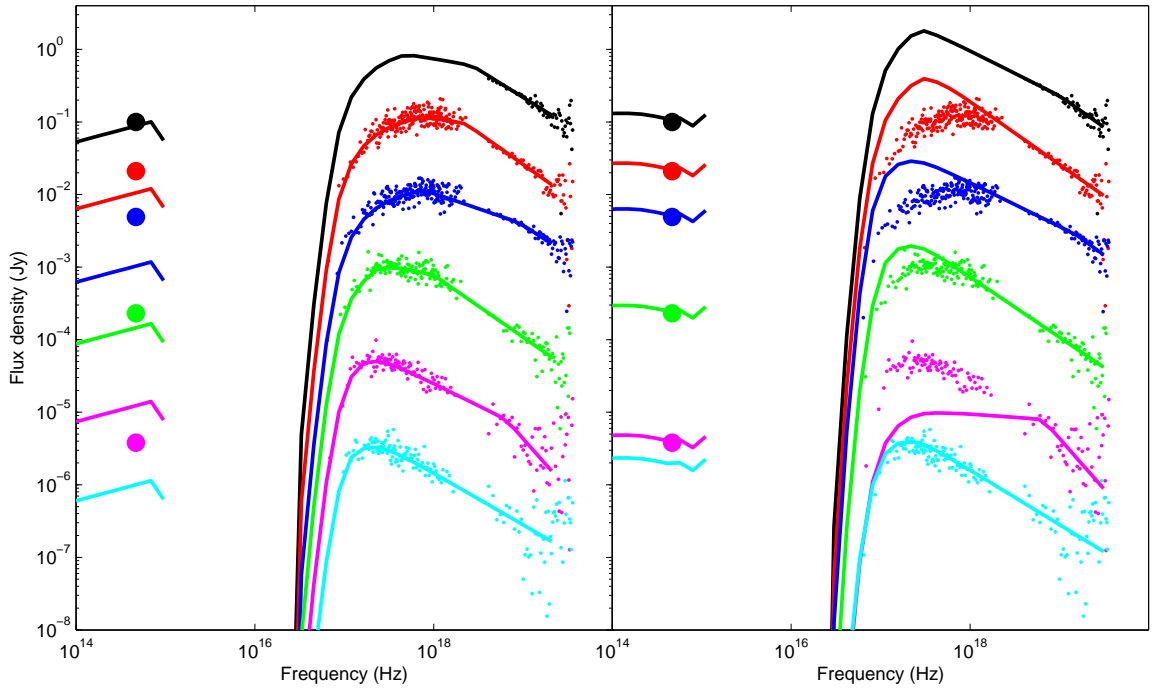}
\caption{\label{fig_sed_prompt}Left Panel: Best fit model (indicated in Table \ref{table_fit_x}) using only the X-ray and the BAT data extrapolated down to the optical band. Right Panel: Best fit model using only the optical and BAT data extrapolated in the X-ray band. For both panels, bins 1 to 6 are drawn from top to bottom (and in black, red, blue, green, purple and cyan respectively) with an arbitrary offset. Error bars have been omitted for clarity in the BAT and XRT bands. The size of the error bars in optical is similar to the symbol size. The optical point 6 has, by construction, a null flux. See the electronic version for colors.}
\end{figure*}

We first note that the extrapolation of our high-energy model (constructed in section \ref{ssec_analysis_HEspec}) severely under predicts the optical flux (Fig. \ref{fig_sed_prompt}), in apparent contradiction with the remarkable correlation of the visible and gamma-ray light-curves. The optical fluxes used here are corrected for the following reverse shock and forward shock emission (see next sections) by subtracting these contributions. Note that according to this, the bin 6 prompt optical flux is equal to zero.

After the detection of GRB~080319B, the "naked-eye" burst, people have studied models that can explain a bright visible emission correlated with the prompt gamma-ray light-curve. For instance \citet{fan09, bes10, has11} have done so, but additional studies will be required to check whether these models apply to \grb\ as well. Making a global fit on the optical and gamma-ray data only, using a broken power law extincted by our galaxy and the host galaxy (using the information obtained from the afterglow SED for the host optical extinction), we obtain a very good fit with $\chi^2_\nu = 1.02$, 335 d.o.f (with the low-energy spectral index fixed to  $=-0.33$), like previously observed in other GRBs \citep{ves05, ves06}. This model is however strongly rejected (with a $\chi^2_\nu = 2.79$), when X-ray data are considered, stressing the importance of keV to sub-keV energies during the prompt phase for the correct interpretation of GRB spectra. More puzzling, in bins 2-4 the optical--gamma-ray model clearly over predicts the X-ray flux and in other cases (bins 5) under predicts it.

The lack of correlation between the XRT and BAT light-curves (Fig. \ref{fig_prompt}) is another puzzling feature of \grb . In the context of synchrotron emission, it could be explained by the synchrotron frequency crossing the energy range of the XRT, this is however an ad'hoc assumption that would not explain {\it how} the optical emission could be correlated to the gamma-ray one, and it is probably more natural to invoke an additional component superimposed on top of the prompt emission (e.g. photospheric emission, refreshed shocks or a two-component jet). 

In short, the internal/external shock model with synchrotron radiation in its simplest version cannot account for the prompt emission of \grb , especially the SED and the lack of correlation between the prompt X-ray and gamma-ray light-curves. Possible ways out of this problem may involve additional radiating regions or additional radiation mechanisms or both. The data at hand do not allow us to distinguish between these possibilities.

We note that it is the combination of the brightness and large duration of this burst which allowed the detection of its prompt emission simultaneously in the optical, X-ray and gamma-ray bands, and the detailed analysis of the SED of the prompt phase. 

Last, as noted by e.g. \citet{gen09}, the prompt optical emission was usually observed as {\it either} a faint erratic emission or a bright and large single flare. It is now possible, due to these observations, to point out that only the faint erratic emission is indeed related to the internal shock (i.e. a prompt signal in the standard fireball framework). The large and bright re-brightening is related to the reverse shock. If this can be generalized to all GRBs, then only the faint signal tracing the prompt gamma-ray light curve should be called the prompt optical emission, the other being the reverse shock optical emission.

\section{Burst geometry, ejecta properties and surrounding medium properties}
\label{sec_ejecta}

\subsection{Presence of a possible jet}
\label{ssec_ejecta_jet}

The late afterglow light curve (during temporal bin 11) presents a steepening seen in X-ray and in optical, with an asymptotical value of $\sim$2.2. In the standard fireball model, this achromatic steepening is the signature of a decelerated jet \citep{rho97}. In such a case, the decay index after the break gives directly the value of p, we thus have p=2.2 $\pm$ 0.2. The date of the jet break, about $61 000$ s after the trigger, is not unusual.

\subsection{Ejecta properties}
\label{ssec_ejecta_times}

The typical times defining the evolution of GRB emission are the duration of the prompt emission, hereafter \tga , the time of deceleration \tdec , and the time of the jet break \tb . \textit{\grb\ is clearly in the thin shell configuration} with \tga $<$\tdec , allowing to separate the prompt GRB from its afterglow and to measure \tga\ and \tdec\ precisely: \tga =120 sec (using the total duration of signal visibility within the BAT and not t$_{90}$), and \tdec =315 sec, both in the source frame. We have also seen in section \ref{ssec_analysis_late} that \tb $\sim$19000 sec, again in the source frame. 

In the standard model the deceleration time (in the source frame) is given by Eq. \ref{eq_tdec} where $E_{52}$ is the energy released by the GRB in units of $10^{52}$ erg, $n_1$ is the density of the surrounding medium in proton/cm$^3$, z is the redshift, $\Gamma_{300}$ is the Lorentz factor normalized to 300. The break time is given by Eq. \ref{eq_tb} \citep[][]{sar99b}, where $\theta$ is the opening angle of the jet in radian. 

\begin{equation}
\label{eq_tdec}
t_{\rm dec} = 2.71~E_{52}^{1/3}~n_1^{-1/3}~\Gamma_{300}^{-8/3} {\rm s}
\end{equation}

\begin{equation}
\label{eq_tb}
t_{\rm b} =  1.04 \times 10^7~E_{52}^{1/3}~n_1^{-1/3}~\theta^{8/3} {\rm s}
\end{equation}

Interestingly, the standard fireball scenario predicts that the ratio of these two times, given by Eq. \ref{eq_ratio}, depends only on the Lorentz factor and on the opening angle of the jet. Considering the values measured for \grb\ (see next sections), we get $ t_{\rm b}/t_{\rm dec} \sim 60$, leading to $(\Gamma_{300}~\theta) \sim 0.016$, suggesting that \grb\ had a small Lorentz factor and a strong beaming ($\Gamma \sim$ 110 and $\theta \sim$ 2.5\deg\ for instance).

\begin{equation}
\label{eq_ratio}
t_{\rm b}/t_{\rm dec} = 3.8 \times 10^6~(\Gamma_{300}~\theta)^{8/3}
\end{equation}

\subsection{The surrounding medium}
\label{ssec_ejecta_medium}

We define the late afterglow as the period lasting from temporal bin 9 to bin 10. In this part, we do observe a "simple" afterglow, which we attribute to the forward shock, and we can use the closure relations \citep{che04, sar98, rho97} to check for the medium geometry \citep[see][for a list of the closure relation used]{gen07}. We used the X-ray and optical data of the temporal bin 9 for this purpose.

The closure relations applied to the optical data are not in agreement with a fast cooling (with $\nu_c < \nu_m$). We thus have the classical ordering $\nu_m < \nu_c$ of the slow cooling. We know from the SED (see Sec. \ref{ssec_analysis_sed}) that a spectral break lies between the optical and the X-ray band. The closure relations imply that this is the cooling break. We thus have during the temporal bin 9:

\begin{equation}
\nu_m \leq \nu_{opt} < \nu_c < \nu_X
\label{ordre_frequence_bin9}
\end{equation}

In such a case, however, X-ray data do not allow to decide between the ISM and the wind medium. The optical data could allow to discriminate them, but due to the large error bars both solutions are compatible with the data within $3\sigma$. Thus, we cannot conclude from the closure relations alone the type of medium surrounding the burst.

In an ISM case, the cooling frequency decreases as $t^{-0.5}$ and would have crossed the X-ray band (assuming the value derived in Sec. \ref{ssec_analysis_sed}) about 50 seconds after the trigger, i.e. before the start of the XRT observation. Conversely, in a wind case, the cooling frequency would start crossing the X-ray band $1.9 \times 10^6$ seconds after the trigger (assuming no jet effect), not observable by the XRT. It is thus impossible to conclude on the surrounding medium. It is however known that usually an ISM medium fit the data better \citep[see e.g.][]{gen07}. In the following, we will use this hypothesis.

\section{Modeling the afterglow}
\label{sec_AGmodel}
The previous considerations set general constraints on the fireball (energy, Lorentz factor, jet opening angle), and we can now try to model the afterglow observations. In the standard afterglow model, only two mechanisms can explain the strong optical re-brightening observed between temporal bins 6 and 8: the start of the afterglow or the reverse shock.

Assuming the re-brightening to be the start of the afterglow, like in \citet{mol07}, the maximum of the emission is emitted at the deceleration radius. In such a case the initial Lorentz factor of the fireball, $\Gamma$ has to be low (of the order of 100) to have a deceleration time in agreement with the peak time. However, the optical decay index of the optical emission after the peak of the afterglow is either 0.25 or $(3p-2)/4$ ($\sim 1.15$ for p=2.2) \citep{sar98, sar99a}. This is not in agreement with the data.

The second solution explains the optical re-brightening with the reverse shock. This is better supported from the data, as the SED during the temporal bin 7 implies the presence of an additional component in optical (see Sec. \ref{ssec_analysis_sed}). According to \citet{kob00}, in the thin shell configuration and slow cooling, the light-curve of the reverse shock has a characteristic evolution, rising like t$^5$, and decaying like t$^{-2}$ (for p=2.2) when the observed frequency falls between $\nu_m$ and $\nu_c$. This is precisely the evolution of the optical light-curve in the interval 300-5000 sec, strengthening the reverse shock interpretation. This interpretation also provides a natural explanation of the small (flatter) transition phase near the bin 8, due to the transition between the reverse and the forward shocks.

We can now use the observed value $t_{dec} = 315$ s in the source frame to estimate the parameters of the jet. If we consider a surrounding medium of constant density n = 0.1 cm$^{-3}$, and an energy release E$_0 = 145 \times 10^{52}$ erg (considering \eiso = $4.34~10^{53}$ erg and assuming 30\% radiating efficiency), we get $\Gamma = 125$, again suggesting a rather low Lorentz factor. 

The situation in the X-ray band during the re-brightening is also complex. We do not expect the reverse shock to be visible in X-rays. However, two features cannot be explained straightforwardly by the forward or the reverse shocks. First, the large spectral variations between the temporal bin 6 to 9, and second a bright X-ray flare peaking at about 600 seconds. For the former discrepancy, we may observe the transition between central engine activity and the classical afterglow. For the latter, we note that other GRBs with a strong delayed optical re-brightening display nearly simultaneous X-ray flares, this is the case of GRB~060418 and GRB~060607A \citep{mol07}, and more recently GRB~100219A\citep{mao11}. Within the framework of the fireball model, X-ray flares superimposed on the afterglow are interpreted as late activity of the central engine. Late central engine activity could also explain the slope of the X-ray spectrum immediately after the flare, $\beta_X = 0.7$ during bin 7, which is equal to the slope measured during the prompt phase (Table \ref{table_fit_x}). Finally the duration of the X-ray flare T$\sim$250 seconds is comparable with the duration \t9\ of the prompt phase, also suggesting that delayed activity of the central engine is continuing at that time, and dominates the X-ray light-curve. One issue with this assumption is that it provides no explanation for the similarity of the temporal decay in X-rays and visible shown in Fig. \ref{fig_alpha}. 

We can then complete our model description. We assume that the emission in bin 9 is completely dominated by the forward shock, that the visible emission in bin 7 is dominated by the reverse shock, and that the X-ray emission in bin 7 is dominated by the prompt. In order to reproduce the observations, we have to consider different microphysics parameters for the reverse shock and forward shock, which is compatible with the fireball model (Mochkovitch, priv. communication). We assume a surrounding medium of constant density n = 0.1 cm$^{-3}$ and an energy release E$_0 = 145 \times 10^{52}$. This leads to jet opening angle of $\theta_j \sim$ 2.1\deg\ \citep{fra01}. The total energy release is thus $4.9 \times 10^{50}$ erg. We take into account the value of the cooling frequency between $10^{15}$ and $10^{16}$ Hz during bin 9, and an early transition from fast to slow cooling. Using the above parameters, the fireball model can reproduce the complete set of data, if we chose p = 2.2 and the following microphysics parameters: $\epsilon_{e,f} = 10^{-2}$ and $\epsilon_{B,f} = 8 \times 10^{-3}$ for the forward shock, and $\epsilon_{e,r} = 10^{-2}$ and $\epsilon_{B,r} =  10^{-1}$ for the reverse shock ($\epsilon_e$ and $\epsilon_B$ being respectively the fractions of energy going into the electrons and the magnetic field). We list the model parameters and its main predictions in Table \ref{table_fit_para}. We stress that this is not a unique fit, and that other sets of parameter values could also reproduce the data, the density in particular is almost a free parameter in our modeling (for instance a similar fit can be obtained for n$_1$ = 10$^{-2}$ cm$^{-3}$, with $\epsilon_{e,f}  = 10^{-2}$ ; $\epsilon_{B,f} = 3 \times 10^{-2}$ for the forward shock, and $\epsilon_{e,r} = 2 \times10^{-2}$ ; $\epsilon_{B,r} = 2 \times 10^{-1}$ for the reverse shock). The values of $\epsilon_B$ for the reverse and forward shocks, lead to $R_B = \sqrt{{\epsilon_{B,r} \over \epsilon_{B,f}}} = 3.5$. $R_B > 1$ is in agreement with the analysis of \citet{gao11}, and with the suggestion of \citet{zha03} that bright optical flashes of reverse shock origin require reverse shocks which are more magnetized than the forward shock. In the following of the paper, we use the values listed in Table \ref{table_fit_para}.

While this paper was in preparation, \citet{cuc11}, \citet{gao11} and \citet{zhe11} published other studies of GRB 110205A. While our results basically agree with those of \citet{zhe11} and \citet{gao11}, they differ from \citet{cuc11} since in our model the optical flux around the maximum of the re-brightening (at t$_{\rm dec}$) is largely dominated by the reverse shock, while we found that it is dominated by the forward shock with the parameters chosen by Cucchiara et al. (2011; these parameters are $\Gamma_0$ = 200 , p = -2.9 , $\nu_{m,f} = 1.2 \times 10^{15}$ Hz , $\nu_{m,r} = \nu_{m,f}/\Gamma_0^2$ , $F_{\nu max,r} = F_{\nu max,f} \times \Gamma_0$, in a slow cooling regime leading to F$_f$/F$_r$ $\sim 30$ at t$_{\rm dec}$ in the R band).

GRB 110205A is one of the few GRBs with a bright visible re-brightening rising after a few hundred seconds. Such bright optical re-brightenings are often attributed to the emission of the shocked ejecta when it encounters the surrounding medium (reverse shock). We can use this observation to discuss the conditions for the existence of bright optical re-brightenings \citep[and their absence in the majority of GRBs,][]{klo09c}. In our model we find that the reverse shock can dominate the visible emission of the forward shock if the microphysics parameter $\epsilon_B$ is larger in the reverse shock. In order to observe a well defined re-brightening, the fireball must also be in thin shell model implying a not too high Lorentz factor. We may thus speculate that GRBs that do not show bright optical re-brightenings rising after the end of the prompt emission are either in the thick shell condition with a high Lorentz factor, or they have similar microphysics parameters in the forward and reverse shocks.

In conclusion our interpretation favors a model in which bin 7 is dominated by the emission from the jet : the reverse shock with a smooth light-curve at optical wavelength and spiky residual prompt activity in X-rays. Bin 9, on the contrary, is dominated by the emission from the forward shock, with a cooling frequency between the optical and X-ray frequencies.

\begin{deluxetable}{ccc}
\tabletypesize{\scriptsize}
\tablecaption{\label{table_fit_para}Parameters of the model that correctly fits the observed light curves and spectra. Predicted values are obtained from \citet{fra01}, \citet{sar98}, \citet{pir04}, and \citet{kob03}}
\tablewidth{0pt}
\tablehead{\colhead{Parameter} & \colhead{Hypothesis value} & \colhead{Predicted value} \\}
\startdata
E$_0$                          & $145 \times 10^{52}$ ergs  & ---                       \\ 
$n_1$                          & 0.1                        & ---                       \\
$p$                            & 2.2                        & ---                       \\ 
$\epsilon_e$ forward shock     & 0.01                      & ---                       \\
$\epsilon_B$ forward shock     & 0.008                      & ---                       \\
$\epsilon_e$ reverse shock     & 0.01                      & ---                       \\
$\epsilon_B$ reverse shock     & 0.1                      & ---                       \\
$\Theta_j$                     & ---                        & 2.1\deg                   \\
E$_\gamma$                     & ---                        & $4.9 \times 10^{50}$  erg \\
$\Gamma$                       & ---                        & 125                       \\
$\nu_m$ bin 9                  & ---                        & $1.3 \times 10^{12}$ Hz   \\
$\nu_c$ bin 9                  & ---                        & $3.9 \times 10^{15}$ Hz   \\
F$_{\nu,max}$ reverse shock (R band) & ---                  & 10.3 mJy                  \\
F$_\nu$ forward shock bin 9 (R band) & ---                  & 0.5 mJy                   \\

\enddata
\end{deluxetable}

\section{Conclusions}
\label{sec_conclu}

We have presented the data of the Swift burst \object{GRB 110205A} taken in optical with several French facilities ranging from 0.25 m up to 1m. This burst is one of the best observed gamma-ray bursts, and quite remarkably, the data show the various radiation features expected in the fireball model:
\begin{itemize}
\item The prompt phase, seen from high energy to the optical band;
\item The reverse chock, seen in optical and well separated from the internal shocks;
\item The classic forward shock, seen in optical, in near-infrared and in X-rays;
\item The jet break and post-jet break light curve.
\end{itemize}

Regarding the interpretation of the very rich data set available on \grb , we tried modeling the observations within the framework of the classic fireball model. Quite surprisingly, the model is able to explain most features of \grb\ without requiring fine tuning of the parameters or additive hypotheses. If "archetype GRBs" exist, GRB~110205A is certainly one of them.  While, we do not have the innocence to believe that the fireball model can explain every single GRB, one of its successes is that it is able to explain so many features of \grb . One issue concerns the interpretation of the prompt emission that is not straightforward within the standard model. We have shown that X-ray data are crucial in the analysis of the prompt spectrum, justifying special efforts to measure the broadband high-energy spectrum of GRBs from below 1 keV to above 1 MeV.

The two missing components in the observing campaign of \grb , namely the supernova signature about 2-3 weeks after the burst and the host galaxy contribution, were too faint to be observable with the instruments we had in hand. This is already one of the lessons GRB 110205A is teaching us: good GRB follow-up must start within seconds with small diameter telescopes, it must continue without interruption for at least a few hours (TAROT lost the rise between the prompt phase and the maximum of the reverse shock) with very good temporal sampling, and it must be followed by a one-month survey on a 4-meter telescope (or larger) dedicated to this work. It would also be interesting to observe the field with large facilities in order to gather informations on the host galaxy of this burst. We already know that the dust model that best fits the data is a Small Magellanic Cloud model. 

During the prompt phase, we have observed a strong correlation between the optical and the BAT light curves. This leads to another of the main conclusion of this paper: we have now reached the limit of the current instrumentation on robotic telescopes. The trailed image is specific to the TAROT project. This feature has shown here (see also \citet{klo08}) its power to study the correlation between optical and high energy light curves. We are however lacking spectral information that would have strongly constrained the nature of the optical emission. Observations with a prism or a similar instrument would help here.

We finally note that a pure Band law cannot fit the high-energy spectral data, which require at least an additional break in the hard X-ray range. Even if the Band law is empirical, it has been fairly good at reproducing the data up to now. This non-agreement is tricky and again should be investigated in line with the other properties of this burst: if we consider GRB 110205A as an archetypal GRB, this observation could indicate that the standard modeling of the prompt emission, with the Band function or simpler functions, is not appropriate to describe the true spectral shape of GRBs, when it is measured from below 1 eV to above 100 keV.

\acknowledgments

We thank R. Mochkovitch and F. Daigne for helpful discussions about the modeling of GRB afterglows and the anonymous referee for insightful suggestions. 
The TAROT telescope has been funded by the {\it Centre National de la Recherche Scientifique} (CNRS), {\it Institut National des Sciences de l'Univers} (INSU) and the Carlsberg Fundation. It has been built with the support of the {\it Division Technique} of INSU. We thank the technical staff contributing to the TAROT project: A. Abchiche, G. Buchholtz, A. Laloge, A. Mayet, M. Merzougui, A.M. Moly, S. Peruchot, H. Pinna, C. Pollas, P. \& Y. Richaud, F. Vachier, A. le Van Suu.
The T1m Telescope is funded by Institut de M\'ecanique C\'eleste et de Calcul des \'Eph\'em\'erides, Observatoire de Paris (IMCCE), Observatoire Midi Pyr\'en\'ee (OMP), Programme National de Plan\'etologie (PNP-INSU).
This work made use of data supplied by the UK Swift Science Data Centre at the University of Leicester. This work has been financially supported by the GdR PCHE and the GDRE "Exploring the Dawn of the Universe with Gamma-Ray Bursts" in France. This work has been partially supported by ASI grants I/009/10/0.

{\it Facilities:} \facility{Swift}, \facility{TAROT}, \facility{OHP-80cm}, \facility{Pic du Midi-T1m}.

\end{document}